\documentclass[12pt]{iopart}

\usepackage{iopams}
\usepackage{graphicx}
\usepackage{xcolor}
\usepackage{hyperref}

\def\ef{\varepsilon_{\mathrm{F}}}
\def\Ec{E_{\mathrm{c}}}
\begin{document}
\title{Coulomb staircase in an asymmetrically coupled quantum dot}

\author{G McArdle$^1$, R Davies$^2$, I V Lerner$^1$, I V Yurkevich$^2$}

\address{$^1$ School of Physics and Astronomy, University of Birmingham, Birmingham, B15 2TT}
\address{$^2$ School of Informatics and Digital Engineering, Aston University, Birmingham, B4 7ET}
\ead{i.v.lerner@bham.ac.uk}

\begin{abstract}
 We investigate the Coulomb blockade in quantum dots asymmetrically coupled to the leads for an arbitrary voltage bias focusing on the regime where electrons do not thermalise during their dwell time in the dot. By solving the quantum kinetic equation, we show that the current-voltage characteristics are crucially dependent on the ratio of the Fermi energy to charging energy on the dot. In the standard regime when the Fermi energy is large, there is a Coulomb staircase which is practically the same as in the thermalised regime. In the opposite case of the large charging energy, we identify a new regime in which only one step is left in the staircase, and we anticipate experimental confirmation of this finding.
\end{abstract}

\submitto{\JPCM}
\maketitle

\section{Introduction}

The phenomenon of the Coulomb blockade in quantum dots has been a longstanding topic of interest and many aspects of it have been studied (see \cite{Aleiner_Review, Alhassid_Review, Kouwenhoven1997} for reviews). It arises due to the strong  Coulomb interaction resulting in large charging energy, $E_{\mathrm{c}} = e^2/C$, that must be overcome in order to add an additional electron onto the dot of capacitance $C$. This leads to a number of notable physical results such as peaks in the conductance as a function of gate voltage \cite{Kulik, AvLik_Paper, Beenakker} and a staircase in the dependence of current on the bias voltage ($I$-$V$ characteristics) that has become known as the Coulomb staircase \cite{Kulik, Averin-Likharev_book_chapter, Ben-Jacob_Wilkins}.

{A prominent approach to understanding transport in mesoscopic systems is based on the classical master equation \cite{AvLik_Paper, Averin-Likharev_book_chapter, Hershfield}, which has typically assumed full thermalisation on the dot. However, a master equation approach is not limited to only dealing with the thermalised case, and the quantum master equation provides a full microscopic description by including the traced-out leads, with the assumption of thermalisation being made to simplify calculations. Using this approach, the full counting statistics of the problem can be calculated under a Markovian approximation \cite{Bagrets_Nazarov, Marcos2010Counting, Li2004RelaxationQubit, Flindt2008Counting}, with recent progress in calculating noise for non-Markovian tunnelling to second order \cite{Xu2022MemoryQmeq}. Other approaches have been successful, such as using the Ambegaokar-Eckern-Sch\"on (AES) action \cite{AES} to study relaxation dynamics on a quantum dot \cite{AES_Relaxation} -  although this method cannot be utilised in all regimes \cite{AES_Problem_Limit}. The non-equilibrium Green's function approach has also been used to highlight the relation between the Coulomb blockade and the zero-bias anomaly \cite{Kamenev_Gefen, Altshuler_Aronov, ZBA_AAL, Alt_Arononv_Book_Chapter}, as well as to calculate the tunnelling density of states of a Coulomb-blockaded quantum dot near equilibrium \cite{Kamenev_Gefen,TDoS}.}

{The assumption of thermalisation is justified when the quasiparticle decay }rate due to the electron-electron interaction, $\gamma $,  is much larger than the tunnelling rates to the (left and right) leads, $\Gamma_{{\mathrm{L,R}}}$, so that the time spent by the extra electrons on the dot is sufficient for their full thermalisation.

In this paper we consider the regime where one can neglect thermalisation,
 \begin{equation}\label{gamma}
    \gamma \ll\Gamma  ,
\end{equation}
otherwise  keeping  the separation of energy scales characteristic for the classical Coulomb blockade \cite{Kouwenhoven1997}:
 \begin{equation}\label{scales}
 \hbar \Gamma \ll \Delta \ll k_B T \ll E_\mathrm{c},
\end{equation}
where $\Delta$ is the typical energy level spacing and $T$ is the temperature. {The rest of this paper will set the Boltzmann and reduced Plank constant to equal one, $\hbar, k_B = 1$.}
The regime (\ref{gamma}) is important,  in particular, when electrons in the dot experience localisation in the Fock space \cite{AKGL} (the precursor for many-body localisation \cite{BAA}) and is easily reachable in metallic quantum dots with a large dimensionless conductance $g$. {We additionally consider the regime where there are a large number of electrons on the dot ($N \gg 1$).} Previously, analytical calculations for this regime have been performed in the linear response limit \cite{Beenakker}, while numerical calculations for an arbitrary bias voltage \cite{Averin_Korotkov} have been limited to the experimentally important regime \cite{Staircase_Expt} when $\ef\gg E_{\mathrm{c}}$ with $\ef$ being the Fermi energy on the dot.
The opposite limit of considerable experimental and theoretical interest is that of a few electrons on the dot, where the lowest energy levels make a strong impact on the observables (see \cite{Few_electron_review} for a review), and the fine structure of the Coulomb staircase is resolved \cite{AltshulerTinkham}.

Here we consider a quantum dot in the absence of thermalisation with strong asymmetry in the coupling to the leads (typically assumed in considerations of the thermalised regime \cite{Kulik, Averin-Likharev_book_chapter, Ben-Jacob_Wilkins, Hershfield}) {for both large and small ratio $\ef/E_{\mathrm{c}}$}. We use the quantum kinetic equation to develop a full analytical solution for the Coulomb staircase for
 $N\gg1$ at any voltage $eV$.

 The solution crucially depends on the ratio $\ef/\Ec$. For $\ef\,{\gg}\,E_{\mathrm{c}}$,  the absence of thermalisation does not play a significant role and the Coulomb staircase remains practically the same as in the thermalised regime \cite{Kulik, Averin-Likharev_book_chapter, Ben-Jacob_Wilkins, Hershfield}, with an equilibrium
established with the most strongly coupled lead.

However, for $\ef\,{\ll}\, E_{\mathrm{c}}$ we show that the staircase practically vanishes. Instead, assuming the traditional anisotropy in coupling to the leads, $\Gamma_{\mathrm{R}}\ll\Gamma_{\mathrm{L}}$, with the voltage $V$ applied to the left lead, there is a single step in the current equal to $e\Gamma_{\mathrm{R}} (N_0{+}1) $ (with $N_0$ being the number of electrons on the dot at $V{=}0$) when  $V$ increases from $0$ to $eV\sim E_{\mathrm{c}}$. All the further steps are of order $1$ in the same units  of $e\Gamma_\mathrm{R}$, i.e.\ practically invisible for $N\gg1$. {
This result is complimented with a numerical calculation using the quantum master equation approach, showing that features of this very strong charging energy regime persist even for $N\lesssim10$.} This is due to a significant contribution of the low energy levels even for a large number of electrons in the dot.

\section{Model}\label{Sec:Model}
We consider the quantum dot asymmetrically coupled to two leads with the bias voltage $V$ applied to the left one described by the Hamiltonian
\begin{equation}\label{H}
    H = H_\mathrm{d} + H_\ell  + H_\mathrm{T}\,.
\end{equation}
Here $H_{\mathrm{d}}$ is the Hamiltonian of the Coulomb-blockaded dot in the zero-dimensional limit \cite{Aleiner_Review, Alhassid_Review,Kouwenhoven1997},
\begin{equation}\label{H_dot}
    H_\mathrm{d} = \sum_n \varepsilon_n d_n^\dagger d_n + {\textstyle{\frac{1}{2}}} {E_\mathrm{c}} \left(\hat{N} - N_\mathrm{g} \right)^2,
\end{equation}
where $\varepsilon_n$ are the energy levels of the dot, $d_n^\dagger \left(d_n\right)$ are the creation (annihilation) operators of the quantum dot, $\hat{N} = \sum_n d_n^\dagger d_n$ is the number operator for the dot, and $N_\mathrm{g}$ is the preferable number of electrons on the dot in equilibrium set by the gate voltage. {The leads are described by}
\begin{equation}\label{H_ell}
    H_\ell  = \sum_{k, \alpha} \left(\varepsilon _{k}-\mu _\alpha\right ) c_{k, \alpha}^\dagger c_{k,\alpha},
\end{equation}
where $\alpha=\mathrm{L}, \mathrm{R}$ labels the lead, $c_{k,\alpha}^\dagger \left(c_{k, \alpha}\right)$ are the creation (annihilation) operators for an electron of energy $\varepsilon _k$, and  $ \mu_\alpha$ is the chemical potential of the lead,  $\mu_\mathrm{L} = \mu +eV$ and $\mu_\mathrm{R} = \mu$. The tunnelling between the dot and the leads is described by the tunnelling Hamiltonian
\begin{equation}\label{H_T}
	H_\mathrm{T} = \sum_{\alpha, k, n} \left( t_{\alpha } c_{k,\alpha}^\dagger d_n + \mathrm{h.c.} \right),
\end{equation}
where the tunnelling amplitude $t_{\alpha }$, which is  assumed to be independent of $k$ and $n$,  defines  the broadening of the energy levels $\Gamma = \Gamma_{\mathrm{L}} + \Gamma_{\mathrm{R}}$ with $\Gamma_\alpha = 2 \pi \nu_\alpha |t_\alpha|^2$, with  the density of states $\nu_\alpha$ taken to be a constant.

We assume the absence of thermalisation in the dot which will allow us to use the quantum kinetic equation for a given energy. This is justified when the inequality  (\ref{gamma})  is satisfied.   For a zero-dimensional diffusive dot, the quasiparticle decay rate due to the electron-electron interaction at energy  $\varepsilon$  is given for $\Delta\ll T$ by \cite{AKGL, Sivan_Imry_1994, Blanter_rates}
\begin{equation}\label{QP_Lifetime}
	\gamma(\varepsilon) \approx \Delta\left(\frac{\varepsilon}{E_\mathrm{Th}}\right)^2,
\end{equation}
where  $E_\mathrm{Th} = g\Delta$ is the Thouless energy and $g \gg 1$ is the dimensionless conductance of the dot. This result is valid provided that $\sqrt{g}\Delta < \varepsilon < E_\mathrm{Th}$.

In the equilibrium regime in the absence of the coupling to the leads, the tunnelling density of states has some interesting features \cite{TDoS}  which, intuitively, are preserved if one lead dominates the behaviour of the system and the chemical potential on the dot will be determined by that lead. This quasi-equilibration allows us to solve exactly the case of strongly asymmetrically coupled leads, either for $\Gamma_{\mathrm{L}}/\Gamma_{\mathrm{R}}\gg1$ when the jumps in the current exist, or for $\Gamma_{\mathrm{L}}/\Gamma_{\mathrm{R}}\ll1$ when the current has almost Ohmic behaviour.

\section{Quantum kinetic equation}\label{Sec:Method}

To analyse the Coulomb blockaded quantum dot in the non-linear regime we use the  Keldysh technique (see, e.g.,  \cite{Rammer_Smith} for a review) in a way similar to that detailed in \cite{Meir_Wingreen_Jauho}.

\subsection{Quantum dot in the weak coupling limit}

In the case of an isolated dot, i.e.\ totally neglecting the level broadening $\Gamma$,  the Keldysh Green's function can be written as a sum over all levels, $g^{>,<}(\varepsilon) = \sum_n g^{>,<}_n(\varepsilon)$ with the single-level Green's functions given by
\begin{equation}\label{GF_Defn}
	g_n^>(t) = -i \Tr\left(\hat{\rho}_0 d_n(t)d_n^\dagger\right), \hspace{5pt} g_n^<(t) = i \Tr\left(\hat{\rho}_0 d_n^\dagger d_n(t)\right),
\end{equation}
where $d_n(t) = \mathrm{e}^{iHt}d_n\mathrm{e}^{-iHt}$ and $\hat{\rho}_0$ is the density matrix. {Additionally,} the particle number is conserved and the Green's functions can be written as sums over the $N$-particle subspaces,
\begin{eqnarray}\label{g>_isolated}
    \fl g^>_n(\varepsilon) = -2\pi i \sum_N \delta \left(\varepsilon - \varepsilon_n - \Omega_N \right) g^>_N(\varepsilon_n), \hspace{10pt} &g^>_N(\varepsilon_n) = \Tr_N \left(\hat{\rho}_0 d_n d_n^\dagger \right), \\
	\label{g<_isolated}
    \fl g^<_n(\varepsilon) = -2\pi i \sum_N \delta \left(\varepsilon - \varepsilon_n - \Omega_{N-1} \right) g^<_N(\varepsilon_n), \hspace{10pt} &g^<_N(\varepsilon_n) = -\Tr_N \left(\hat{\rho}_0 d_n^\dagger d_n \right),
\end{eqnarray}
with the normalisation $\sum_N\left(g^>_N(\varepsilon_n) - g^<_N(\varepsilon_n) \right) = 1$. The charging energy required to add an electron is included above through  $\Omega_N$ defined as
 \begin{equation}\label{OmegaN}
    \Omega_N\equiv  E_{N+1}-E_N = E_{\mathrm{c}}\Big(N+{\textstyle{\frac{1}{2}}}-N_{\mathrm{g}}\Big),\qquad E_N \equiv {\textstyle{\frac{1}{2}}} E_\mathrm{c} (N-N_\mathrm{g})^2 .
 \end{equation}

 The coupling to the leads is included via the quantum kinetic equation (QKE), which in the weak coupling limit $(\Gamma \rightarrow 0)$ can be written for each level as \cite{Meir_Wingreen_Jauho,Haug_Jauho}
\begin{equation}\label{QKE_original}
	g^{>, <}_n \left(\varepsilon\right) = g^{\mathrm{R}}_n \left(\varepsilon\right) \Sigma^{>, <} \left(\varepsilon\right) g^{\mathrm{A}}_n \left(\varepsilon\right).
\end{equation}
The self energies for non-interacting leads are assumed to be independent of the dot level $n$ and are given by
\begin{eqnarray}\label{Self-energy>}
    \Sigma^>(\varepsilon) &=& \sum_{k, \alpha} |t_\alpha|^2 g^>_{k,\alpha}(\varepsilon) = - i \left[\Gamma - \left(\Gamma_{\mathrm{L}} f_{\mathrm{L}}(\varepsilon) + \Gamma_{\mathrm{R}} f_{\mathrm{R}}(\varepsilon) \right) \right],
\\\label{Self-energy<}
    \Sigma^<(\varepsilon) &=& \sum_{k, \alpha} |t_\alpha|^2 g^<_{k,\alpha}(\varepsilon) = i\left(\Gamma_{\mathrm{L}} f_{\mathrm{L}}(\varepsilon) + \Gamma_{\mathrm{R}} f_{\mathrm{R}}(\varepsilon)\right).
\end{eqnarray}
Above, the Green's functions for the leads are $g^>_{k,\alpha}(\varepsilon) = -2\pi i (1-f(\varepsilon-\mu_\alpha)) \delta(\varepsilon - \varepsilon _k+\mu _\alpha )$ and $g^<_{k,\alpha}(\varepsilon) = 2\pi i f(\varepsilon-\mu_\alpha) \delta(\varepsilon - \varepsilon _k+\mu _\alpha )$, where $f(\varepsilon-\mu_\alpha)$ is a Fermi function. The density of states in the leads, which enters via the tunnelling rates $\Gamma_\alpha = 2 \pi \nu_\alpha |t_\alpha|^2$, is given by $\nu_\alpha = \sum_k \delta(\varepsilon - \varepsilon _k+\mu _\alpha )$, while $\Gamma = \Gamma_{\mathrm{L}} + \Gamma_{\mathrm{R}}$. {Note that the form of (\ref{QKE_original}), with all functions being considered at the same energy, corresponds to no thermalisation with $\gamma\rightarrow0$. This rate must be the smallest scale in the system for the hierarchy of scales in (\ref{gamma}, \ref{scales}) to be satisfied, therefore it can be taken to zero with no issues. }

Now we rewrite the QKE (\ref{QKE_original}) as
\begin{equation}\label{QKE_Simple}
    g_n^>(\varepsilon)\Sigma^<(\varepsilon) = g_n^<(\varepsilon) \Sigma^>(\varepsilon).
\end{equation}
Substituting in Eqs.~(\ref{g>_isolated}, \ref{g<_isolated}) we use the  ansatz
 \begin{equation}\label{ansatz}
    g^>_N(\varepsilon_n) = p_N\left(1-F_N(\varepsilon_n)\right)\quad \mathrm{ and }\quad g^<_N(\varepsilon_n) = -p_N F_N(\varepsilon_n),
 \end{equation}
where  $p_N$ is the probability of having $N$ electrons on the dot and $F_N(\varepsilon_n)$ is the distribution function given $N$ electrons on the dot which, in the case of complete thermalisation, goes over to the equilibrium Fermi distribution function. In these terms, we write
  the QKE as follows:
\begin{eqnarray}\label{QKE}
    p_N \left(1-F_N(\varepsilon_n)\right)\widetilde {f}(\varepsilon_n + &\Omega_N) &= p_{N+1}F_{N+1}(\varepsilon_n) \left(1-\widetilde {f}(\varepsilon_n + \Omega_N) \right),
\end{eqnarray}
where
\begin{eqnarray}\label{ftilde}\widetilde {f}(\varepsilon) &= \frac{\Gamma_{\mathrm{L}}}{\Gamma}f (\varepsilon-\mu -eV) + \frac{\Gamma_{\mathrm{R}}}{\Gamma}f(\varepsilon-\mu ).
\end{eqnarray}
 This corresponds to the detailed balance equations derived in \cite{Beenakker} for $\Delta \gg T$ and reproduces the case of complete thermalisation after the summation over $n$ and making the replacement $F_N (\varepsilon )\to f(\varepsilon-\ef )$.
 The QKE  (\ref{QKE}) should be complemented by the normalisation conditions, $\sum_{N} p_N=1$ and $\sum_{n} F_N(\varepsilon _n)=N$.

We represent the current going from the dot to the lead $\alpha$  via $p_N$ and $F_N(\varepsilon_n)$ as
\begin{equation}\label{I_lead}
	\fl I_\alpha = e\Gamma_\alpha \sum_N p_N \sum_n \Big(F_N(\varepsilon_n)\left[1-f (\varepsilon_n {-}\mu _\alpha {+} \Omega_{N{-}1})\right]- \left[1-F_N(\varepsilon_n)\right] f (\varepsilon_n {-}\mu _\alpha {+} \Omega_N) \Big)
\end{equation}
Applying current conservation, $ I=I_{\mathrm{R}}=-I_{\mathrm{L}} $  and using $\mu _{\mathrm{L}}=\mu +eV$ and $\mu _{\mathrm{R}}=\mu $, we express the current  as
\begin{eqnarray}\label{Current}
	\fl I = e\frac{\Gamma_{\mathrm{L}} \Gamma_{\mathrm{R}}}{\Gamma} \sum_N p_N \sum_n &\Big( F_N(\varepsilon_n)\left[f (\varepsilon_n -\mu_{N-1} -eV) - f (\varepsilon_n-\mu_{N-1})\right] \nonumber\\&+ (1-F_N(\varepsilon_n)) \left[f(\varepsilon_n -\mu_N -eV) - f (\varepsilon_n -\mu_N) \right] \Big).
\end{eqnarray}
with $\mu_N \equiv  \mu-\Omega_N$. Assuming
 a  density of states on the dot to be constant, $1/\Delta$, we convert the sum over $n$ to an integral over all energies on the dot (counted from zero). Then in the low-$T$ limit
\begin{equation}\label{Current_Modified}
\fl
I = e\frac{\Gamma_\mathrm{L}\Gamma_\mathrm{R}}{\Gamma}\sum_N p_N \Bigg[ \int_{\mu_{N-1}}^{\mu_{N-1} +eV} \mathrm{d}\varepsilon \,\Theta(\varepsilon) F_N(\varepsilon) +  \int_{\mu_N}^{\mu_N+eV} \mathrm{d}\varepsilon \,\Theta(\varepsilon) (1-F_N(\varepsilon))\Bigg],
\end{equation}
where $\Theta(\varepsilon)$ is the Heaviside step function.

\subsection{Solution to the QKE}
The charging energy strongly penalises states with a wrong number of electrons on the dot.  In the case of strongly asymmetric leads with $\Gamma_{\mathrm{L}}\gg\Gamma_{\mathrm{R}}$, the main contribution to (\ref{Current}) is given by the two states with $N$ closest to $N_{\mathrm{g}}+eV/E_{\mathrm{c}}$, since  electrons have time to fill the dot up. In the opposite case,   $\Gamma_{\mathrm{L}}\ll\Gamma_{\mathrm{R}}$,  the two relevant states are those closest to $N_{\mathrm{g}}$. Keeping only the appropriate two states in the QKE (\ref{QKE})  allows us to obtain the following exact solution:
\begin{eqnarray}\nonumber
	p_N &= \frac{Z_N}{Z_N+Z_{N+1}}, \qquad \qquad p_{N+1} &= \frac{Z_{N+1}}{Z_N+Z_{N+1}}, \\[-9pt]\label{Full_Soln}\\[-9pt]
	F_N(\varepsilon_n) &= \frac{Z_N(\varepsilon_n)}{Z_N}, \qquad \qquad F_{N+1}(\varepsilon_n) &= \frac{Z_{N+1}(\varepsilon_n)}{Z_{N+1}}, \nonumber
\end{eqnarray}
where
\begin{eqnarray}
 	Z_N &= \sum_{\{n_j=0,1\}} \prod_{j=1}^\infty \left[\varphi(\varepsilon_j+\Omega_N)\right]^{n_j} \delta_{(\sum_j n_j), N},
\nonumber \\[-9pt]\label{Z_defn} \\[-9pt]\nonumber
 	Z_{N+1} &= \sum_{\{n_j=0,1\}}  \prod_{j=1}^\infty \left[\varphi(\varepsilon_j+\Omega_N)\right]^{n_j} \delta_{(\sum_j n_j), N+1},
\end{eqnarray}
with functions $\varphi $ defined via $\widetilde{f}$ in (\ref{ftilde})  as
\begin{equation}\label{varphi}
	\varphi(\varepsilon_j + \Omega_N) = \frac{\widetilde {f}(\varepsilon_j + \Omega_N)}{1 - \widetilde {f}(\varepsilon_j + \Omega_N)},
\end{equation}
 while $Z_{N}(\varepsilon_n)$ in (\ref{Full_Soln}) is defined by restricting the
   sums in (\ref{Z_defn})  to  configurations with the state  $\varepsilon_n$ occupied.  It is important to highlight that due to the form of the QKE (\ref{QKE}), $Z_{N+1}$ in (\ref{Z_defn}) contain $\Omega_N$ rather than $\Omega_{N+1}$ so that the relevant $N$ dependence enters only in the Kr\"{o}necker delta.

When $N\gg1$, the Kr\"{o}necker delta is equivalent to a delta function,
 \begin{equation}\label{Eq:Delta_n_sum}
	\delta_{(\sum_j n_j), N} = \int \frac{\mathrm{d}\theta}{2\pi}\mathrm{e}^{i\theta\left(\sum_j n_j - N\right)}\,,
\end{equation}
which allows us to write the sums in (\ref{Z_defn}) in the form
\begin{equation}\label{Z_saddle}
	Z_N = \int \frac{\mathrm{d}\theta}{2\pi} \mathrm{e}^{Nf(\theta)}, \qquad f(\theta) = \frac{1}{N}\sum_j \ln\left( 1+ \varphi(\varepsilon_j+\Omega_N)\mathrm{e}^{i\theta}\right) -i\theta.
\end{equation}
{Now $Z_N$ can be evaluated in the saddle-point approximation. The optimal $\theta_0$ is found from the second equation above where the sum is converted to the integral, $\sum_{j} \to \Delta^{-1}\int_{0}^{\infty}{\mathrm{d}}\varepsilon$, which gives}
\begin{equation}\label{Saddle_Point}
{\ef}=N{\Delta} =
\int_0^\infty {\mathrm{d}\varepsilon}\left({\frac{\mathrm{e}^{-i\theta_0}}{\varphi(\varepsilon + \Omega_N)}+1}\right)^{-1}.
\end{equation}
As $\Omega_N$ is unchanged by definition when going between $Z_N$ and $Z_{N+1}$, (\ref{Z_defn}),
  the relevant $N$ dependence of $\theta_0$ enters only via {$\ef=N\Delta$. Thus we find that in the saddle-point approximation  $Z_N = g(\theta_0)\mathrm{e}^{-iN\theta_0}$, where $g(\theta_0)$ is a function which depends on $N$ only via $\ef$.  Hence for $N\gg1$,  this function} is approximately the same  for $Z_N$ and $Z_{N+1}$ which allows us to cancel $g(\theta_0)$ in calculating $ {p_N}$ and  $F_N(\varepsilon_n)$  in (\ref{Full_Soln}). This results in
\begin{equation}\label{p and F}
	\frac{p_{N+1}}{p_N} = \mathrm{e}^{-i\theta_0}, \hspace{10pt} F_N(\varepsilon_n)   \approx F_{N+1}(\varepsilon_n) \approx \left({\frac{\mathrm{e}^{-i\theta_0}}{\varphi(\varepsilon + \Omega_N)}+1}\right)^{-1}.
\end{equation}
 The ratio of probabilities can be found by using $N = \sum_n F_N(\varepsilon_n)$, which corresponds to the saddle point equation above.

 The resulting $I$-$V$ characteristics turn out to be strikingly different for the two opposite regimes,  when the ratio $\varepsilon_\mathrm{F}/E_\mathrm{c}$ is either small or large, as described in the following section.

\section{Results and Discussion}\label{Sec:Results}
We begin by reproducing the well-known results of the standard theory for $\ef\gg E_{\mathrm{c}}$ to show that (i) our approach works and (ii) the absence of the full thermalisation does not make a significant impact on the Coulomb staircase in the case of strong asymmetry in the coupling to the leads.

Then we show that in the opposite limit, $\ef\ll E_{\mathrm{c}}$, there is only one significant step left in the Coulomb staircase if $N\gg1$. Additionally, we present numerical results for small $N$ which are in full agreement with our analytical results for $N\gg1$.

\subsection{Small charging energy, $E_\mathrm{c}\ll \ef  $}
We start with the linear response regime. Then $\widetilde f (\varepsilon )\to f(\varepsilon -\mu )$ in (\ref{ftilde}) so that $\varphi(\varepsilon+\Omega_N) \to \exp[-\beta(\varepsilon-\mu+\Omega_N)]$ in (\ref{varphi}). Hence, using (\ref{p and F}) we reduce the saddle point equation (\ref{Saddle_Point})  to
\begin{equation}\label{Saddle_Eqm}
	\ef = \int_0^\infty \frac{\mathrm{d}\varepsilon}{\mathrm{e}^{\beta(\varepsilon-\mu_N)-i\theta_0}+1} = T\ln\left(\mathrm{e}^{\beta\mu_N+i\theta_0} + 1\right) \approx \mu_N+i\theta_0T,
\end{equation}
where the approximate equality holds in the low-temperature limit, $\beta\mu_N +i\theta_0 \gg 1$. {The result in (\ref{Saddle_Eqm}) leads to $i\theta_0=\beta(\ef-\mu_N)=\beta (\ef -\mu +\Omega_N)$ (with $\mu $ being the chemical potential in the leads and $\ef$ in the dot), meaning that the low-temperature limit corresponds to $\beta\varepsilon_\mathrm{F} \gg 1$ satisfying the conditions in (\ref{scales}). Furthermore, substituting  into (\ref{p and F}) the expression for $i\theta_0$,} and using $ p_N+p_{N+1} \approx 1$ results in the following expressions for the probabilities and distribution function,
\begin{equation}\label{Eqm_Results}
	p_N = \frac{\mathrm{e}^{-\beta(E_N+N(\ef-\mu))}}{\sum_N \mathrm{e}^{-\beta(E_N+N(\ef-\mu))}}, \hspace{10pt} F_N(\varepsilon) = \frac{1}{\mathrm{e}^{\beta(\varepsilon-\ef)}+1},
\end{equation}
where the sum over $N$ is restricted to the two states with $N$ closest to $N_{\mathrm{g}}$.
Substituting (\ref{Eqm_Results}) into the current (\ref{Current}) results in the following shape of the differential conductance near the peak, $\mu-\Omega_N-\ef = 0$:
\begin{equation}\label{Peak_Shape}
	G = \frac{\mathrm{d}I}{\mathrm{d}V} = \frac{e^2}{2\Delta} \frac{\Gamma_{\mathrm{L}}\Gamma_{\mathrm{R}}}{\Gamma} \frac{\frac{\beta}{2}(\Omega_N+\ef-\mu)}{\sinh(\frac{\beta}{2}(\Omega_N+\ef-\mu))},
\end{equation}
in agreement with \cite{Kulik, Beenakker}.

We now turn to the nonlinear regime and demonstrate, by reproducing the well-known results \cite{Kulik, Averin-Likharev_book_chapter,Ben-Jacob_Wilkins} for strongly asymmetric coupling to the leads and $\ef\gg E_{\mathrm{c}}$, that  the absence of thermalisation has no impact   on the Coulomb staircase.
For $\Gamma_{\mathrm{L}} \gg \Gamma_{\mathrm{R}}$, the solution to  the QKE (\ref{QKE}) for any $V$ is given by (\ref{Eqm_Results}) provided that we replace  $\mu $ by $\mu _{\mathrm{L}}\equiv \mu+eV $   and  restrict the sum over $N$    to the two states with $N$ closest to $N_{\mathrm{g}}+eV/E_{\mathrm{c}}$. Due to the exponential forms of the probabilities in (\ref{Eqm_Results}), only one such state contributes to the current outside some narrow windows in $V$. For a given $V$, this is the state where $N$ obeys the inequality $\Omega_{N-1} \lesssim eV \lesssim \Omega_N$. Noticing that the distribution function in  (\ref{Eqm_Results}), $F_N(\varepsilon) = f(\varepsilon-\ef)$, is a Fermi function with a chemical potential $\ef $, we see that the second integral in (\ref{Current_Modified}) does not contribute to the current for low $T$, as the  upper limit of integration $\mu_N+eV\approx\ef-(\Omega_N-eV)<\ef$.

\begin{figure}[b]
	\mathindent=0pt
	\includegraphics[width = 0.45 \textwidth]{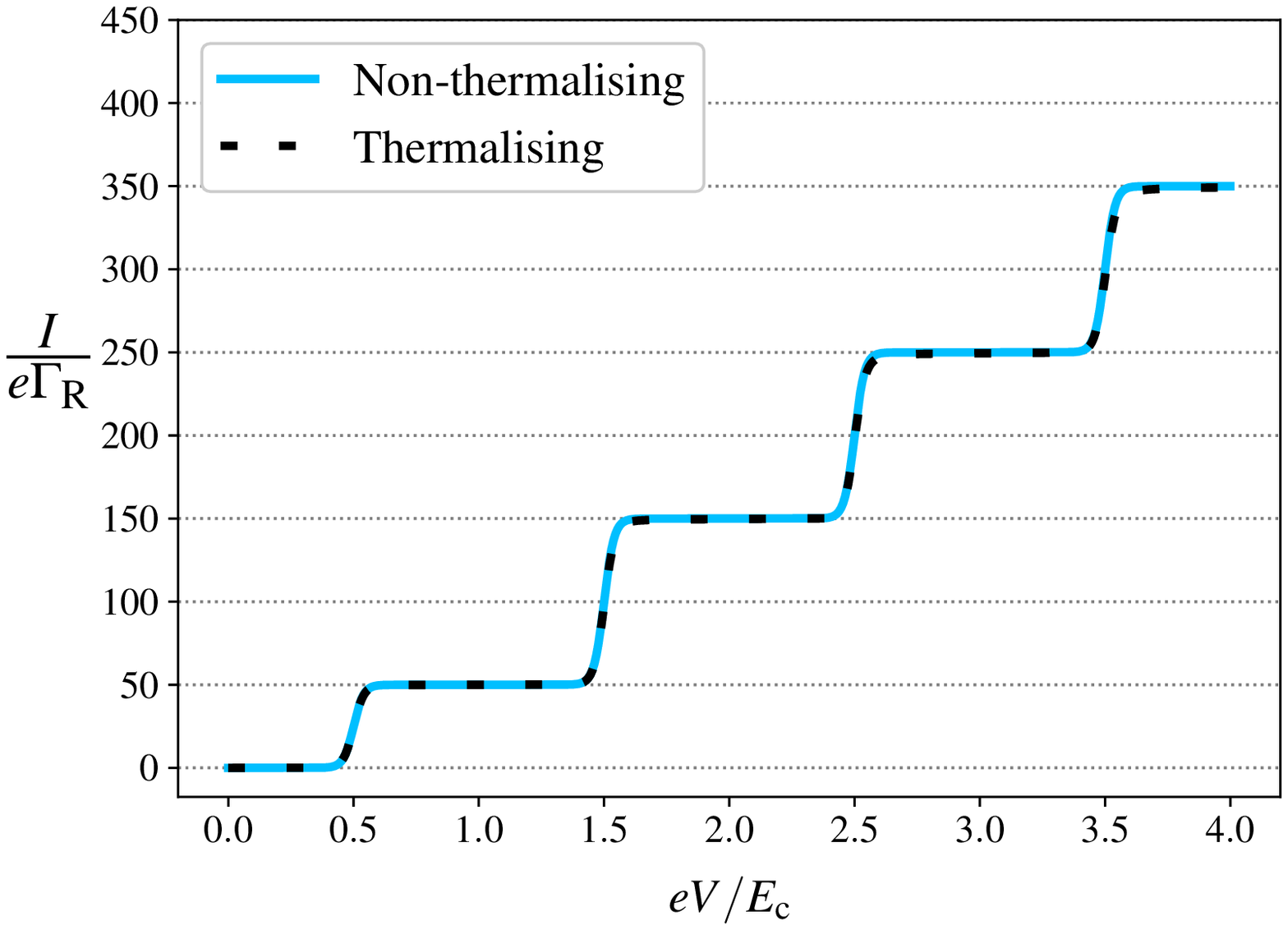}\qquad\includegraphics[width = 0.45 \textwidth]{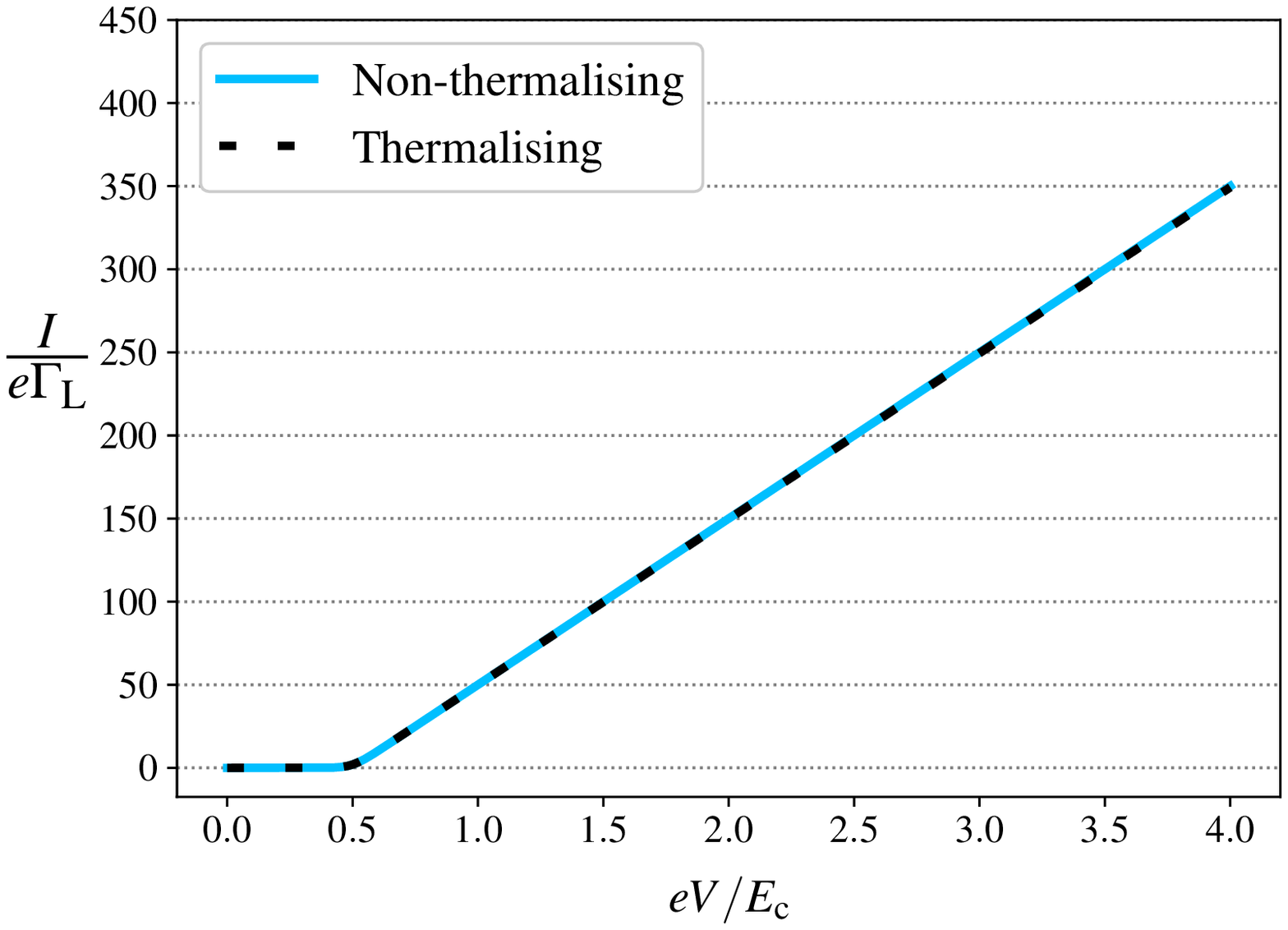}
{\hspace*{.22\textwidth}\textbf{({a})}\hspace*{.47\textwidth}\textbf{({b})}}
	\caption{\label{fig:smallEc}The $I$-$V$ characteristics for a dot in the regime where $N_0\Delta\gg E_\mathrm{c}$ ($N_0\Delta {=} 10 E_\mathrm{c}$) and $\Omega_{N_0} = E_\mathrm{c}/2$. The blue line represents our solution to the QKE and the black dashed line is the solution to the master equation in the standard theory where full thermalisation is  {assumed \cite{Kulik, Averin-Likharev_book_chapter, Ben-Jacob_Wilkins}.} In both instances, \textbf{(a)}: $\Gamma_{\mathrm{L}}/ \Gamma_{\mathrm{R}} = 10^3$ and \textbf{(b)}: $\Gamma_{\mathrm{L}}/ \Gamma_{\mathrm{R}} = 10^{-3}$ an equilibrium is set up with the dominant lead and the approaches produce the same results.}
\end{figure}

Consider the contribution of the first integral in (\ref{Current_Modified}), starting with the regime that begins in equilibrium ($V=0$) and continues for $0\le eV \lesssim \Omega_{N_0}$, when there are $N_0$ electrons on the dot.  Then, as $\Omega_{N_0} \approx E_\mathrm{c}/2$  the lower integration limit  {$\mu _{N_0-1}\equiv \mu -\Omega_{N_0-1} \approx\ef+E_\mathrm{c}/2>\ef$ so that this integral also vanishes}.  The current is therefore zero as expected. With $V$ increasing beyond  $\Omega_{N_0}$, there are $N>N_0$ electrons on the dot. In this case, having $\ef \gg E_\mathrm{c}$ ensures that $\ef > \Omega_N$ for all relevant $\Omega_N$ and both the integration limits are positive, so the presence of $\Theta(\varepsilon )$ is irrelevant. The steps in the current  in the low-$T$ limit are, therefore, given by
\begin{eqnarray}
	I = 0, & \qquad 0 &\leq eV \lesssim \Omega_{N_0} \qquad \quad   (p_{N_0} = 1), \nonumber
\\[4pt]
	I = e\Gamma_{\mathrm{R}} \frac{\Omega_{N_0}  }{\Delta} , & \quad \Omega_{{N_0}}   &\lesssim eV\lesssim \Omega_{{N_0}+1}  \qquad  (p_{{N_0}+1} = 1),\label{Large_Results}\\[4pt]
	 I = e\Gamma_{\mathrm{R}} \frac{\Omega_{{N_0}+1}  }{\Delta} ,\qquad &\Omega_{{N_0}+1}   &\lesssim eV\lesssim \Omega_{{N_0}+2}     \qquad(p_{{N_0}+2} = 1), \nonumber
 \end{eqnarray}
and so on.
This demonstrates a staircase structure with the steps separated by   $eV {=} E_\mathrm{c}$  and an
{almost} constant height proportional to $E_\mathrm{c}/\Delta$. The full results, including the windows around the jumps at $eV{=}\Omega_N$, are obtained by substituting (\ref{Eqm_Results}) with the change $\mu \rightarrow \mu_\mathrm{L}$ into (\ref{Current}) and are practically indistinguishable from the full thermalisation case \cite{Kulik, Averin-Likharev_book_chapter, Ben-Jacob_Wilkins}, as shown in Figure~\ref{fig:smallEc}(a).

For the opposite asymmetry, $\Gamma_{\mathrm{R}} \gg \Gamma_{\mathrm{L}}$, equilibrium with the right lead (with no voltage applied there) is maintained and no staircase is observed as $p_{N_0} \approx 1$ for all values of $V$. Instead, the Ohmic behaviour prevails for $eV\gtrsim \Omega_{N_0}$ as the tunnelling electron gains more energy as shown in Figure \ref{fig:smallEc}(b).

\subsection{Large charging energy, $E_\mathrm{c}\gg \ef $}
In this limit, the low-energy states in the dot make a considerable impact on the transport behaviour. The reason is that the regime  {$ \ef < \Omega_N $,  which was impossible  $\ef/\Ec\gg1$}, now arises.

\begin{figure}[b]
	\mathindent=0pt
	\includegraphics[width = 0.45 \textwidth]{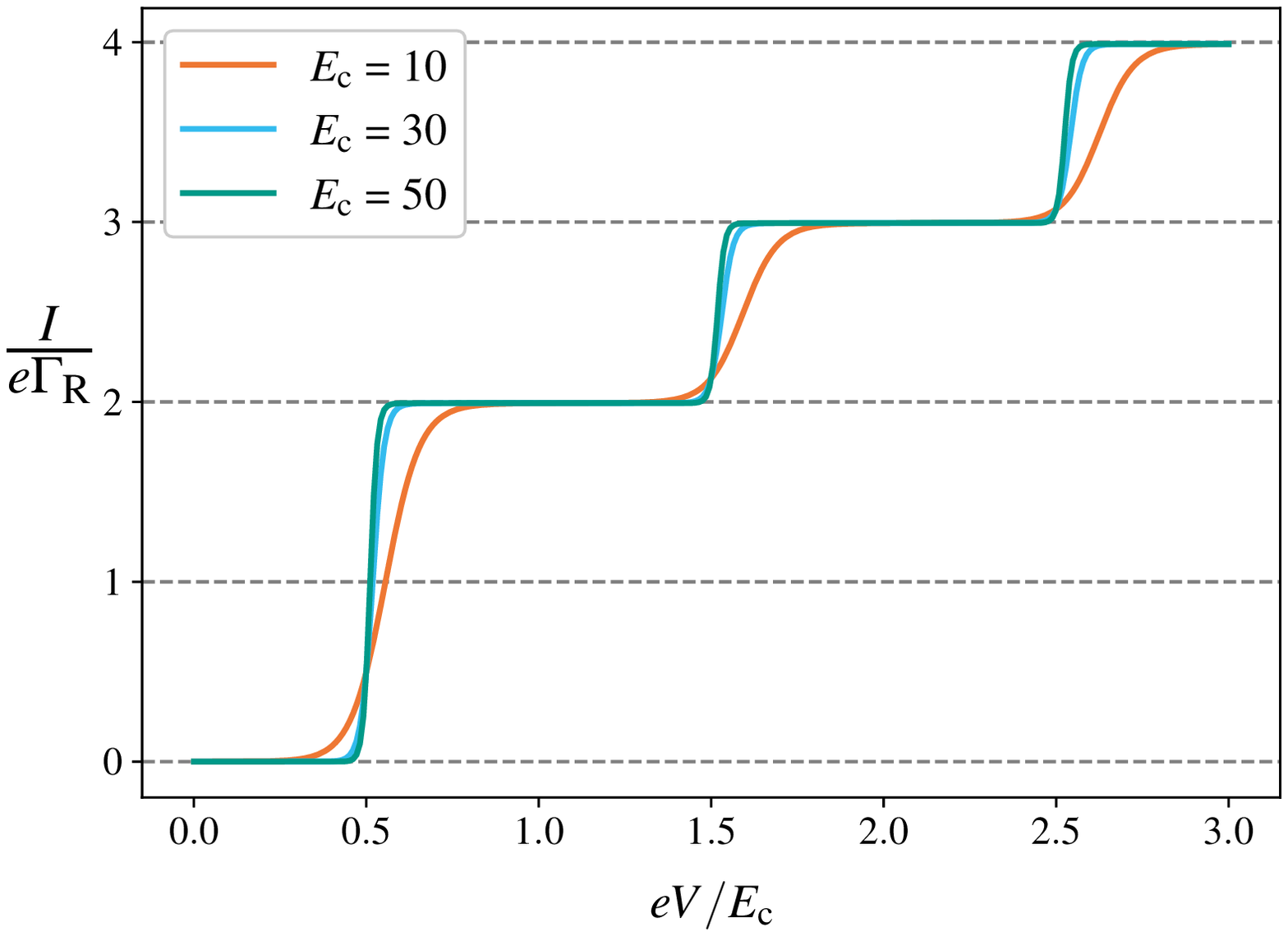}\qquad\includegraphics[width = 0.45 \textwidth]{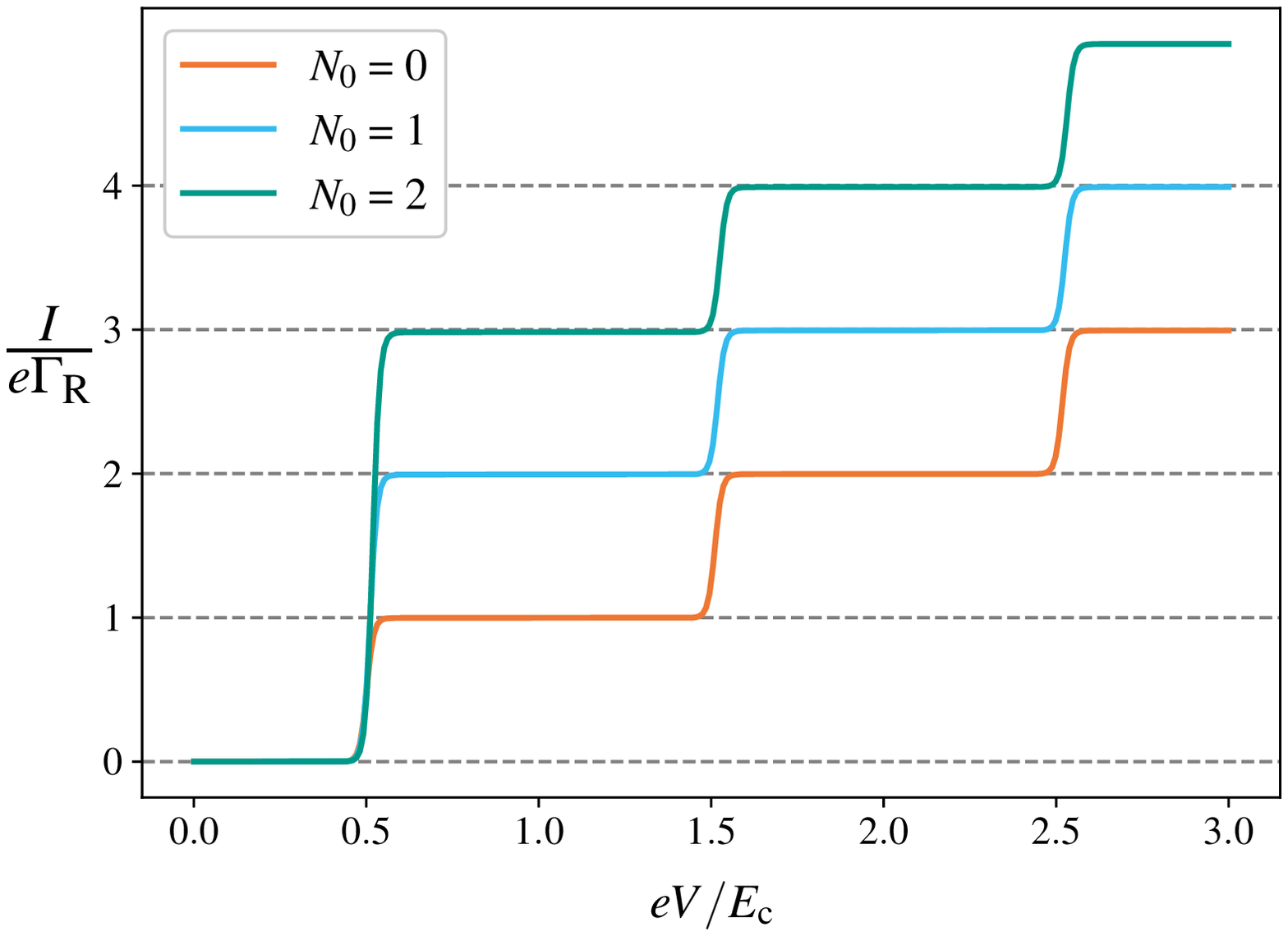}
	\hspace*{.22\textwidth}\textbf{({a})}\hspace*{.47\textwidth}\textbf{({b})}
	\caption{\label{fig:largeEc} The numerical $I$-$V$ characteristics for a dot with 7 states in the regime where $N_0\Delta\ll E_\mathrm{c}$ ($N_0\Delta\approx 0.01 E_\mathrm{c}$) and $\Omega_{N_0} = E_\mathrm{c}/2$. \textbf{(a)}: Increasing the charging energy makes the steps sharper but does not affect the size of the jumps. \textbf{(b)}: Increasing the number, $N_0$, of electrons in equilibrium (with the gate voltage)  illustrates that the first jump is equal to $e\Gamma_R(N_0{+}1)$. In both cases $\Gamma_\mathrm{L} = 100\Gamma_\mathrm{R}$.}
\end{figure}

 For $\Gamma_\mathrm{L}\gg\Gamma_\mathrm{R}$,  the expressions for $p_N$ and $F_N(\varepsilon )$ are formally the same as for $E_\mathrm{c}\ll \ef$ in (\ref{Eqm_Results}) with the substitution $\mu \rightarrow \mu_\mathrm{L}$. However, as $F_N(\varepsilon )$ is now an extremely narrow function (on the scale of $E_{\mathrm{c}}$) and the integration limits may be negative, the contributions of the above integrals to the current are severely restricted in comparison to the case of $\ef/\Ec\gg1$.
Starting again with $N_0 $ electrons on the dot at equilibrium, we make similar arguments as in the former case to see that only the first integral in (\ref{Current_Modified}) contributes. The crucial difference  for $N > N_0$ is that the lower limit of integration,  $\mu _{N-1}\approx \ef-\Omega_{N-1}=N\Delta-\Omega_{N-1} $, is less than zero, so that $\Theta (\varepsilon )$ becomes relevant. Therefore, we find the current in the low-$T$ limit to be strikingly different from that in (\ref{Large_Results}). (Note that for the opposite asymmetry,   $\Gamma_{\mathrm{R}} \gg \Gamma_{\mathrm{L}}$, the current remains Ohmic for any ratio $\ef/E_{\mathrm{c}}$.)
\begin{eqnarray}
	I = 0, & \qquad 0 &\leq eV \lesssim \Omega_{N_0} \qquad \quad   (p_{N_0} = 1), \nonumber
\\
\label{Small_Results}I = e\Gamma_{\mathrm{R}} ({N_0}+1),   & \quad \Omega_{{N_0}}   &\lesssim eV\lesssim \Omega_{{N_0}+1}  \qquad  (p_{{N_0}+1} = 1),\\
	I = e\Gamma_{\mathrm{R}} ({N_0}+2)   ,\qquad &\Omega_{{N_0}+1}   &\lesssim eV\lesssim \Omega_{{N_0}+2}     \qquad(p_{{N_0}+2} = 1), \nonumber
 \end{eqnarray}%
and so on.
Crucially the first jump in the current (measured in units of $e\Gamma_{\mathrm{R}}$) at $eV=\Omega_{N_0}$ is equal to  {$N_0+1$} while all the subsequent jumps equal to $1$ in these units.

For $N_0\gg1$, this means that the staircase practically disappears beyond the first step in contrast to the constant jumps of size $E_{\mathrm{c}}/\Delta$ for large $N_0\Delta$, see (\ref{Large_Results}).
Although we have performed analytical calculations for $N_0\gg1$, the results for $\ef \ll E_{\mathrm{c}}$ turn out to be exactly the same for small $N_0$ given a constant charging energy. We demonstrate this by numerically solving the quantum master equation \cite{QmeqPackage} under the  {conditions (\ref{gamma}, \ref{scales})}, for a dot with 7 levels. {This was achieved by solving the first order von Neumann equation for a dot that has $N$ energy levels separated by $\Delta$; the first order equation is sufficient due to the small coupling to the leads. The many-body states on the diagonal of the density matrix are all the $2^N$ occupations with the appropriate charging energy, $E_c(N-N_g)^2/2$, added for the occupation of the configuration. There is no dissipation mechanism for a state to decay on the dot, with relaxation occurring after tunnelling into the leads, therefore the numerical calculations are for the case of zero thermalisation on the dot.} The results are shown in Figure \ref{fig:largeEc}. While all the steps there are pronounced, all but the first one would practically disappear for $N_0\gg1$.

\section{Conclusion}\label{Sec:Discussion}

 To summarise, we have analytically calculated $I$-$V$ characteristics of the quantum dot with a strong asymmetry in the tunnelling coupling to the leads in the Coulomb blockade regime (\ref{scales}) in the absence of thermalisation (\ref{gamma}). We have solved the appropriate quantum kinetic equation in the two limits, for either a large or small ratio, $\Ec/\ef$, of the charging energy to the Fermi energy of electrons in the dot.

 We have demonstrated that for a relatively small charging energy, $\Ec/\ef\ll1$, the absence of thermalisation in a quantum dot has practically no impact on  the Coulomb staircase as an equilibrium is established between the dot and the most strongly coupled lead,  see Figure  \ref{fig:smallEc}. This is in agreement with previous numerical results \cite{Averin_Korotkov} which assume the distribution function is the same for all relevant $N$. We have verified this assumption in the large $N$ limit when no more than two states are relevant   in (\ref{p and F}).

In the opposite limit,  $\Ec/\ef\gg1$, we have analytically shown that for $N\gg1$ the Coulomb staircase has only one pronounced step. With a voltage $V$ applied to the left lead and $\Gamma_{\mathrm{L}}/\Gamma_{\mathrm{R}}\gg1$, this is a step in the current  from $0$ to $e\Gamma_{\mathrm{R}}(N_0+1)$ in a narrow window around $eV=\Omega_{N_0}$ with $\Omega_{N_0}= \Ec/2$ if $N_0=N_{\mathrm{g}}$, see (\ref{OmegaN}). All the subsequent current jumps with $V$ increasing have the magnitude $e\Gamma_{\mathrm{R}}$, see (\ref{Small_Results}), i.e.\ negligible when the number of electrons at equilibrium $N_0\gg1$. Further to the analytic results, we have  numerically solved the quantum master equation for a constant $\Ec$ to find that the analytical results (\ref{Small_Results}) proven for $N\gg 1$ are exactly valid also in the experimentally attractive regime of $N \lesssim 10$, see Figure \ref{fig:largeEc}.  The reason for such behaviour of the Coulomb staircase is that the only electrons available for tunnelling are those in an energy window $\sim\ef$ with the voltage window being much larger, $eV \sim E_\mathrm{c}$. With $\ef/\Ec$ increasing, more electrons are available for tunnelling, thus restoring the jumps between the steps to their full value $\propto \Ec/\Delta$ in the usual regime $\ef\gg\Ec$ \cite{Kulik, Averin-Likharev_book_chapter, Ben-Jacob_Wilkins} where electrons from the entire voltage window contribute to the current.
\section*{Acknowledgements}
We gratefully acknowledge  support from EPSRC  under the grant EP/R029075/1 (IVL) and   from the Leverhulme Trust under the grant  RPG-2019-317 (IVY).

\section*{References}

\end{document}